# Structural and emission properties of $Tb^{3+}$-doped nitrogen-rich silicon oxynitride films

C. Labbé,[1] Y.-T. An,[1] G. Zatryb,[2] X. Portier,[1] A. Podhorodecki,[2] P. Marie [1], C. Frilay [1], J. Cardin [1] and F. Gourbilleau[1]

[1]CIMAP CIMAP, Normandie Univ, ENSICAEN, UNICAEN, CEA, CNRS, 6 Boulevard Maréchal Juin 14050 Caen Cedex 4, France [2]Department of Experimental Physics, Wroclaw University of Technology, 50-370 Wroclaw, Poland

E-mail: christophe.labbe@ensicaen.fr

**Abstract**

Terbium doped silicon oxynitride host matrix is suitable for various applications such as light emitter compatible with CMOS technology or frequency converter system for photovoltaic cells. In this paper, amorphous $Tb^{3+}$ ions doped nitrogen rich silicon oxynitride (NRSON) thin films were fabricated by reactive magnetron co-sputtering method, using various $N_2$ flows and annealing conditions, in order to study their structural and emission properties.

The Rutherford Backscattering Scattering (RBS) measurements and the refractive index values, confirmed the silicon oxynitride nature of the films. An electron microscopy analysis conducted for different annealing temperatures ($T_A$) was also performed up to 1200 °C. The Transmission Electron Microscopy (TEM) images revealed two different sublayers. The top layer showed porosities coming from a degassing of the oxygen element during the deposition and the annealing, while in the region close to the substrate, a multilayer-like structure of $SiO_2$ and $Si_3N_4$ phases appeared, involving a spinodal decomposition. Upon a 1200 °C annealing treatment, an important density of Tb clusters was detected, indicating a higher thermal threshold of rare earth (RE) clusterization in comparison to the silicon oxide matrix.

With an opposite variation of the $N_2$ flow during the deposition, the nitrogen excess parameter ($N_{ex}$) estimated by RBS measurements was introduced to investigate the Fourier Transform

Infrared (FTIR) spectra behavior and the emission properties. Different vibration modes of the Si-N and Si-O bonds have been carefully identified from the FTIR spectra, characterizing such host matrix, especially the "out-of-phase" stretching vibration mode of the Si-O bond.

The highest $Tb^{3+}$ photoluminescence (PL) intensity was obtained by optimizing the N incorporation and the annealing conditions. In addition, according to these conditions, the integrated PL intensity variation confirmed that the silicon nitride-based host matrix had a higher thermal threshold of rare earth clusterization than its silicon oxide counterpart. The analysis of time-resolved PL versus $T_A$, showed the impact of the Tb clustering on the decay times, in agreement with the TEM observations. Finally, the PL and PL excitation (PLE) experiments and comparison of the related spectra between undoped and Tb-doped samples were carried out to investigate the impact of the band tails on the excitation mechanism of $Tb^{3+}$ ions.

## 1. Introduction

The $Tb^{3+}$ ions are attractive energy acceptors for silicon photonics applications due to their important emission lines at blue and green colors ascribed to its $^5D_3$ and $^5D_4$ energy levels. Consequently, $Tb^{3+}$ ions doped silicon oxide films have been investigated as a promising luminescent material compatible with CMOS technology. If we restrict the deposition techniques used to form such a rare earth (RE) doped thin films to implantation, Molecular Beam Epitaxy (MBE), magnetron sputtering and Plasma-Enhanced Chemical Vapor Deposition (PECVD), the most common host matrix for the $Tb^{3+}$ ions is $SiO_2$ [1-4]. Some works in the literature reached a step forward by fabricating various devices based on this matrix [4-7]. Some investigations have also been carried out on $Tb^{3+}$: SiC(N) materials [8-10] as well as on the silicon rich silicon oxide films ($Tb^{3+}$:SRSO)[11] with CMOS devices [12]. However, the incorporation of silicon excess in SRSO, contributing to the formation of Si nanoclusters, significantly reduced the emission intensity of lanthanide ions as a result of a strong non-radiative recombination [13-15].

For this reason, some researchers have focused their works on silicon (oxy)nitride thin films with the aim to find efficient light sources directly integrated on optical chips [16]. Indeed, the silicon oxynitride host is an attractive choice, because of its relatively small band gap (4〜5 eV) which may play an important role in the energy transfer from matrix to RE ions with an efficient electrical injection [17]. Moreover, this matrix should have much higher RE ions solubility in comparison to its silicon oxide counterpart and then allow to prevent RE clustering ions [18]. Furthermore, Si rich silicon nitride (SRSN) has been investigated since Si excess can form clusters which act as luminescence centers to enhance RE emission [19]. In contrast to this, various studies mention the role played by the localized states in the band tails of the amorphous matrix acting as sensitizers for RE emission [16, 20-22]. As underlined by Yerci *et al.*, the addition of N atoms introduces more disorder than in the case of SRSO matrix[16]. In this case, by contrast with the Si excess, the N excess ($N_{ex}$) would introduce many defects leading to the appearance of band tails. For this reason, one can expect an efficient sensitization of RE ions in nitrogen rich silicon (oxy)nitride (NRS(O)N) films [17, 23]. Indeed, for such matrices an intense $Tb^{3+}$ photoluminescence (PL) signal was observed by Jeong *et al.* [24], Yuan *et al.* [25] and by our group [26], even for non-resonant excitation. From these results, two significant applications have been demonstrated using a Si-based host matrix: a CMOS device based on a $Tb^{3+}$:$SiO_xN_y$/$SiO_2$ superlattice [27] and a photovoltaic application based on a down converter system [28, 29]. Therefore, the $Tb^{3+}$ excitation mechanism in NRSON requires further investigations in order to fully understand this key mechanism for applications.

In this study, we investigate the structural and optical properties of Tb-doped NRSON films. These films were fabricated by reactive magnetron co-sputtering method for different reactive nitrogen flow and the annealing temperature ($T_A$). Rutherford Backscattering Scattering (RBS), Transmission Electron microscopy (TEM) as well as Energy-Filtered TEM (EFTEM) measurements have been performed to analyze the composition and structure of these samples. An investigation of different vibration modes of the Si-N and Si-O bonds was conducted using Fourier Transform Infrared (FTIR) spectrometer. Taking into account both the reactive flux and the annealing effect, we reported here the optimal conditions for preparation of highly luminescent samples and discussed the Tb-clusterization effect. Finally, the energy transfer mechanisms between the NRSON host matrix and the $Tb^{3+}$ ions were considered.

**2. Experimental details**

In this work, Tb-doped NRSON films were deposited onto p-type 250 µm-thick (100)-oriented Si wafers by reactive radio frequency magnetron co-sputtering of a pure Si target topped with 5 $Tb_4O_7$ pellets under pure nitrogen plasma. The N gas flux was varied from 3 to 9 sccm, while the plasma total pressure was concomitantly changed in the range of 4-28 µbar. The deposition temperature ($T_d$) and RF power density were fixed at 200°C and 1.23 W.cm$^{-2}$, respectively. The annealing was processed under a given N flow, at temperatures ($T_A$) varying from 500 °C to 1200 °C during 1 hour.

The chemical composition of the layers was obtained by RBS measurements using a 1.5 MeV $^4He^+$ ion beam with a normal incidence and a scattering angle of 165 °, revealing the relative atomic concentrations (% at.) of the different elements. TEM was used to observe the evolution of the structure of the samples upon annealing. For such observations, cross sectional thin foils were prepared by means of a FEI HELIOS Nanolab 660 focused ion beam until electron transparency. The samples were then investigated using a double corrected JEOL ARM200 operated at 200 keV, equipped with a cold field emission gun, and a GATAN imaging filter (GIF) (QUANTUM 965ER spectrometer). The chemical analyses were performed by EFTEM using the image mode of the GIF to analyze the composition variation of the layers. The digitized images were processed by the commercial GATAN software called DIGITAL-MICROGRAPH. The microscope is also equipped with an Energy Dispersive X-Ray (EDX) spectrometer for chemical analyses.

The FTIR spectra were recorded in the range of 500-4000 cm$^{-1}$ using a Nicolet Nexus spectrometer under normal and Brewster's incidence (65 °) angles. The optical properties of the layers have been characterized using Ellipsometry and Photoluminescence spectroscopy measurements. For the former, they were conducted by means of UVISEL Jobin-Yvon ellipsometer with an incident angle of 66.2 °. The experimental spectra were recorded on a 1.5–5 eV range with a 0.01eV resolution. The refractive indexes, given at 1.95 eV, as well as the thicknesses, were deduced from the experimental data by a dispersion law derived from the Forouhi–Bloomer model for amorphous semiconductors using the DeltaPsi2 software [30]. The thickness and refractive index uncertainties are ±10 nm and ±0.01 respectively. Finally, photoluminescence (PL) and PL excitation (PLE) spectra were performed at room temperature by means of a Jobin-Yvon Fluorolog spectrophotometer using a 450W Xe lamp as excitation source with a spectral resolution lower than 5 nm. The PL lifetime was obtained by means of an optical parametric oscillator with a 5 ns pulse at full width at half maximum and a 10 Hz repetition rate for a 244 nm wavelength excitation. The spot diameter was 500 μm with an average energy of 15 mJ.

## 3. *Elemental composition analysis*

Elemental compositions of the as-deposited (AD) samples were determined by RBS measurements. Figure 1(a) shows RBS spectra ranging from 0.20 up to 1.10 keV for the samples obtained with different N flows. The backscattering signals from N and O atoms can be seen in figure 1(a), along with the Si one coming from both thin film and substrate as reported in previous works [14, 27]. Note that the oxygen content in the films is originating from the $Tb_4O_7$ chips. The slight change in the curve shape indicates a small variation of the composition of these films (Figure 1(a)). Figure 1(b) shows RBS spectra in the 1.15 up to 1.40 keV range, which corresponds to the backscattering from $Tb^{3+}$ ions. The width of the Tb-related band changes from sample to sample due to their different thicknesses (230-280 nm), coming from the different N gas flux. The band intensities (RBS yields) are very similar indicating that the Tb content is almost the same for all the films. Moreover, Tb spatial distribution in each sample is quite uniform across the film thickness.

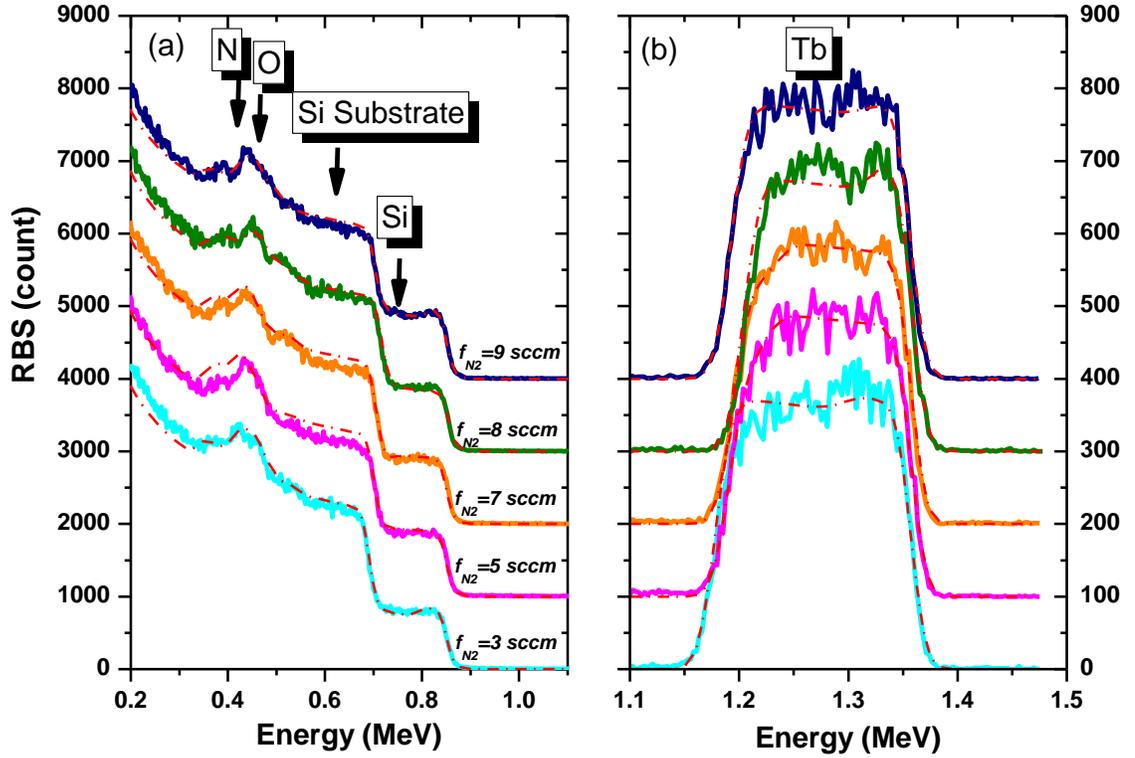

Figure 1. RBS experimental spectra of the AD samples with the corresponding N flow. Si, O and N elements are displayed on the left side (a), whereas the RBS spectra of Tb atoms are shown on the right (b). A simulated curve using SIMNRA software is added to each experimental RBS spectrum (red curves). The spectra have been arbitrarily up-shifted for clarity.

By fitting RBS spectra as shown in figure 1, the film compositions were obtained (see figure 2(a)). The fits require the use of two layers: one at the *top* of the film (so called "*top layer*") in contact with air and another one at the *interface* near the substrate (so called "*interface layer*"). The ratio between the resulting top and interface layer's thicknesses varied from three (190 nm/60 nm) to one third (60 nm/190 nm). Atomic concentrations are calculated considering the weight of the interface and top layers thicknesses. As can be seen in figure 2(a), the Tb concentration is constant in all of the investigated films and equal to 0.73±0.05 at.%. The fact that Tb has a homogenous distribution within the whole film for the AD samples differs from Tb-doped SRSO layers. Indeed, for SRSO host matrix, it has been shown that Tb ions tend to agglomerate and form clusters close to the film/substrate and film/air interfaces, even for low concentration (0.1 at.%.) [14]. As expected, the nitride matrix seems to prevent such an agglomeration process even for 7 times higher Tb concentrations. This issue is discussed further in another section (§7).

Regarding the other elements, the O content varies between 13 at.%, and 17 at.%, while the Si content increases from 29 at.% up to 36 at.%. Surprisingly, a slight decrease of the N concentration of almost 5 at.% is observed with increasing the N flow.

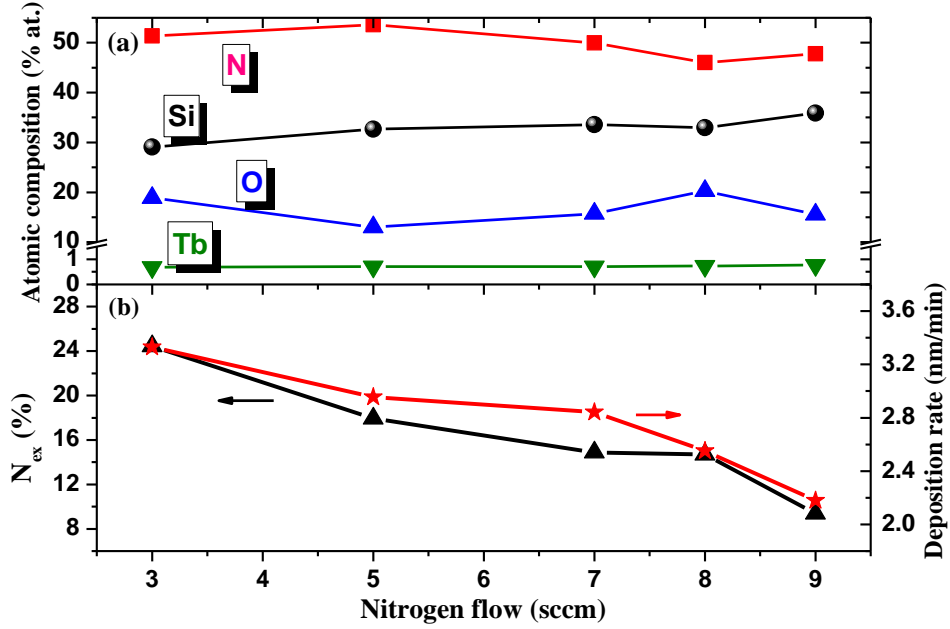

Figure 2. (a) Atomic compositions of the Tb-doped NRSON films as a function of $N_2$ flow (b) $N_{ex}$ (left scale) and deposition rate (right scale) versus $N_2$ flow for AD layers.

To push further investigation, N excess ($N_{ex}$) has been deduced using the following relation:

$$N_{ex}(\%) = \left( \frac{[N] - \frac{4}{3}[Si] + \frac{2}{3}[O] - \frac{7}{6}[Tb]}{[N] + [O] + [Si] + [Tb]} \right) \times 100$$

where the term [element] represents the atomic concentration (at.%) of the element (see supplementary data). This ratio gives the deviation from a perfect $SiO_2$ and $Si_3N_4$ mixture [28, 31] and highlights the N importance compared to the other elements. The N excess evolution as a function of N flow is displayed in figure 2(b). Clearly, $N_{ex}$ decreases with N flow, having values from 24.4 %, 17.9 %, 14.9 %, 14.7 %, to 9.4 % (see the left scale of figure 2(b)) which corresponds to the samples called $S_{24}$, $S_{18}$, $S_{15}$, $S_{14}$, and $S_9$, respectively. Such a behavior seems to be linked with the layer deposition rate (right scale of figure 2(b)), which gradually decreases when the $N_2$ flow increases. Similar result has been obtained by Xu *et al.* [32], who suggested that, from target to substrate, the ejected species collide and/or react with $N_2$ leading to the modification of the mean free path. Thus, some species do not have enough energy to reach the substrate. Consequently, one can assume that a high $N_2$ flow is unfavorable to the incorporation of N atoms and to the deposition rate of the growing film.

## 4. *Refractive index analysis*

The refractive index was also investigated. The bilayer model used to estimate the refractive index by fitting with DeltaPsi software is consistent with the bilayer model used in RBS analysis. The refractive index increases from 1.54 up to 1.79 with N content. These low index values are between those of the stoichiometric $Si_3N_4$ (n=2.04) and $SiO_2$ (n=1.45) materials, confirming that we have fabricated an oxynitride matrix [17, 33, 34].

## 5. *Microscopic analysis of the annealing temperature effect*

In this section, we present different TEM images of the $S_9$ sample in a cross-section configuration for different $T_A$ values. These images were obtained for AD sample and for samples annealed at 700 °C, 1000 °C and 1200 °C respectively, during 1h under N flow (figures 3-5).

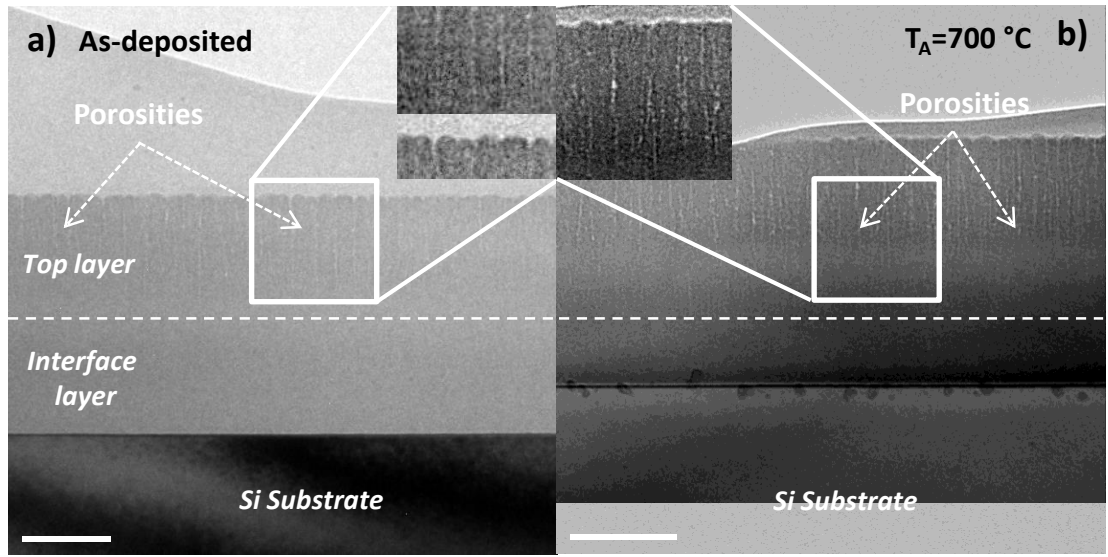

Figure 3: Cross-sectional TEM images for AD (a) and 700°C annealed $S_9$ samples (b). Some vertical porosities are clearly observed in the enlarged images in both cases.

According to these observations, the deposited thin films remain amorphous (no diffracting particles) whatever the post annealing treatment applied. The AD $S_9$ sample presents porosities coming from almost the middle of the film (dashed horizontal line in figure 3(a)) and form vertical tortuous columns of nanoscale diameters in the direction of the top surface (figure 3(a), zoom inset) using a slight defocus condition. These hollow columns are more visible upon 700 °C annealing (figure 3(b), zoom inset) which is probably related to the degassing of light elements from the film (O and/or N) during annealing. But the deposition process itself, occurring almost for two hours at a substrate temperature of 200 °C, seems to have the same effect since similar microstructure is observed for the AD sample. Apparently, this degassing process coming from almost the middle of the film, results in a two layers structure. These results confirm our previous assumption with two modeled layers, named *top* and *interface* layers, used for the RBS and the refractive index modeling. Note that this limit between these layers can change without apparent logic between one quarter to three-quarters of the full thickness. The same kind of degassing has been also confirmed by TEM observations on other samples (not shown here).

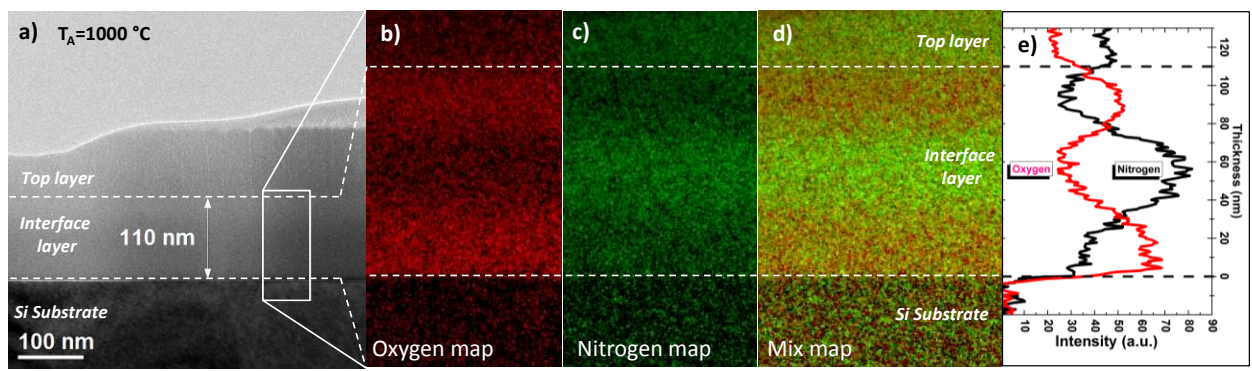

Figure 4. 1000 °C annealed $S_9$ sample investigated through: (a) a cross-sectional bright field TEM image, the corresponding elemental EFTEM maps of (b) oxygen in red color, (c) nitrogen in green color, (d) mix map and (e) chemical profiles of the N and O from the region delimited by a white square in the TEM image. The ratio of integrated profiles between N and O amounts corresponds to the ratio of the RBS measurements (34.39 at.% for N and 30.59 at.% for O) inside the interface layer.

To investigate this issue in more detail, the $S_9$ sample has been annealed at higher temperature ($T_A$=1000 °C) and EFTEM images (figure 4(a)), displaying O and N maps, have been recorded for the whole thickness of this film, by using their respective K edges at 532 eV and 401 eV. Unfortunately, the rare earth element concentration remains too low to be imaged by this technique. Noteworthy, the silicon profile map (not shown) has a homogenous distribution in the film, by contrast with the O and N profiles shown in figures 4(b)-(e). Indeed, concerning the interface layer, O is highly concentrated both on the top and on the bottom (figure 4(b)), whereas N is concentrated mostly in the middle (figure 4(c)). Surprisingly, the profiles of these two last elements offer a clear complementary contrast, displaying a kind of multilayer structure. Note that such N, O distributions are also noticed for the AD and all the annealed layers.

Concerning the top layer, the EFTEM maps reveal poor O content while the N is homogeneously distributed as we can see in figure 4(e), where only the bottom of the top layer is visible but the whole top layer has a uniform distribution of these elements. The low O content, confirmed by RBS measurements (0.1 at. %), suggests the degassing coming from this element. The oxygen may diffuse towards the surface, symbolized by the boundary between the top and the interface layer (figures 4(a)-(e)), featuring the possible physical limit for a degassing inside a solid phase. This frontier position does not appear to vary significantly with the thermal budget (figures 3(b) and 4(a)), suggesting that this effect appears during the deposition (figure 3(a)).

This limit as well as the separation of the two different elements (figure 4(d)), highlighting an unmixing inside the interface layer, are strictly parallel to the substrate surface. These boundaries could be induced by the stress evolution of the film during the deposition [35]. More precisely, this observation suggests a stress-induced diffusion of N by the Si substrate [36, 37], or at least an evidence of the significant role of this interface, leading to a phase separation with a predominant $SiO_2$ phase when O content is important and a main $Si_3N_4$ phase if it is weak. Then, the film/substrate interface could be at the origin of a surface-directed spinodal decomposition (SDSD)[38] in the pseudobinary $(SiO_2)_x(Si_3N_4)_{1-x}$ alloy system. As a matter of fact, such unintended multilayer structure has already been observed by Lui *et al.* [39] for $(HfO_2)_x(SiO_2)_{1-x}$ alloy system near the substrate and has been attributed to a spinodal decomposition (SD) [38]. Another similar effect has been noticed for Si-rich-$HfO_2$ materials, but in this case, coming from the air-film interface with $HfO_2$ and $SiO_2$ phases [40]. Such a SDSD process has also been shown in $ZrO_2$-$SiO_2$, $La_2O_3$-$SiO_2$ and $Y_2O_3$-$SiO_2$ systems [41]. The SD is also the process at the origin of the formation of Si nanoclusters in $SiO_x$ or $SiN_x$ host matrices due to the high Si content [42-44]. In our case, the $N_{ex}$ could be at the origin of such effect assisted by a stress-induced diffusion. In addition, the deposition occurs at 200 °C which is a relatively high deposition temperature for a magnetron co-sputtering process, providing a degassing affecting also the SDSD. Indeed, usually, such SD effect generates a composition wave which has a 4-5 nm periodicity. In our case, this periodicity is approximately one order magnitude higher (30-40 nm), impacted by the degassing and the deposition of our thicker film (230-280 nm). In other studies with Si excess reported in the literature, one can find again the periodicity on a large thickness of almost 40 nm, similarly to our case [42, 43].

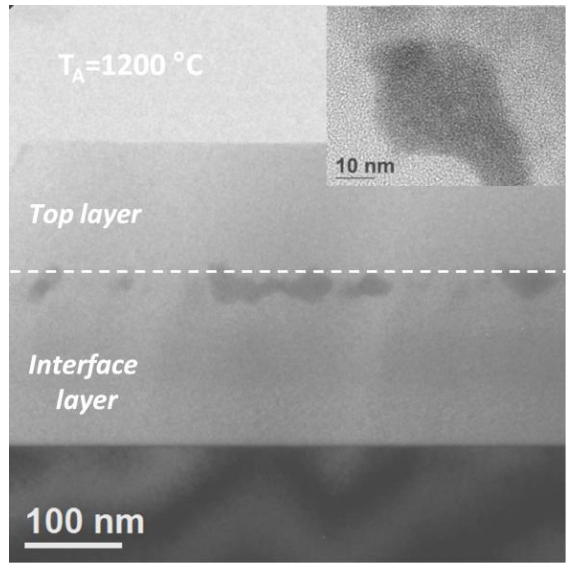

Figure 5. Cross-sectional TEM image of the 1200 °C annealed $S_9$ sample. Tb-rich clusters are visible in the center of the deposited layer. Inset: zoom of a Tb agglomeration.

A TEM image shown in figure 5 was obtained on the $S_9$ sample annealed at 1200 °C, only to check the thermal budget threshold at which clusterization occurs. The elemental EFTEM map (not shown) is commented in the supplementary data. The TEM image reveals an important density of clusters with a darker contrast located below the limit between the top/interface layer (dashed line in figure 5), with a size of about 20-40 nm and a lengthening following this boundary. Note that such clusters have started to be detected (weakly contrasted) for samples annealed at 700°C throughout the thickness of the interface layer. These agglomerates at 1200°C were identified by EDX measurements as amorphous Tb-rich clusters (inset of figure 5). Similar particles have already been observed and identified as Tb-oxide clusters ($TbO_2$, $Tb_2O_3$) in $SiO_2$ by Nazarov *et al.*. Their presence are due to a more stable thermodynamic state of this phase and located in the center of the film for a $T_A$ at 900°C and at the air/film and film/substrate interfaces for a higher $T_A$ [6]. Due to the fact that our clusters stay amorphous, they should not be in the same stoichiometry as mentioned above. In contrast to this observation, a Tb-doped silicon-rich silicon oxide with 50 at.% Si, shows a Tb diffusion from the both film interfaces towards the middle of the film, to become uniform across the film depth [13, 14]. In our case, the surprising position of this clusters line is probably due to the separation between the interface and the top layer, offering a stable energy for the formation of Tb clusters. In any cases, such Tb clusters are completely formed at 1200°C confirming that such host nitride-based matrix is less favorable to RE clustering than its silicon oxide counterpart due to a slower diffusion of RE [18, 27, 45, 46]. In others words, the silicon oxynitride matrix have a higher thermal threshold of RE clusterization compared to the silicon oxide matrix.

## 6. *Infrared spectroscopic analyses*

6.1. Constant Nitrogen excess

FTIR spectra were performed on as-deposited $S_9$ (Figure 6) and well as on all NRSON samples (figure 7). The spectra for the $S_9$ sample were measured under a normal incident angle (figure 6(a)) and Brewster angle (figure 6(b)). As it can be seen, a broad band ranging from 650 cm$^{-1}$ up to 1350 cm$^{-1}$ with no sharp peaks is appearing.

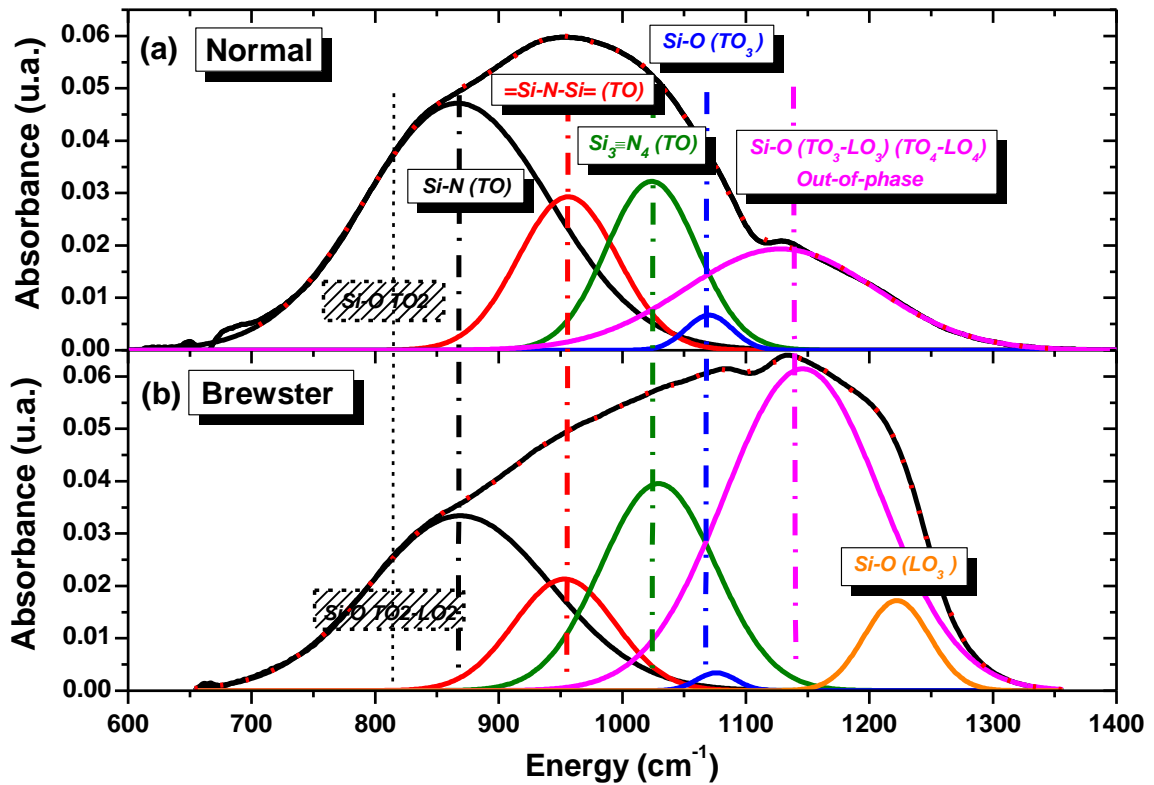

Figure 6. FTIR spectra recorded at normal (a) and Brewster angle (65 °) (b) for $S_9$ AD film where the film absorbance has been normalized to its thickness. The Si-O (LO$_2$) at 810 cm$^{-1}$ and Si-O (TO$_2$) at 820 cm$^{-1}$ are not fitted [47].

To identify the different peaks constituting such a broad band, we have fitted the normal incident angle spectrum with 5 Gaussian peaks (less would not be consistent with the literature cited in table 1), revealing the transversal optical modes (TO) (figure 6(a)). However, the Brewster angle spectrum reveals the longitudinal modes (LO) and therefore an additional 6$^{th}$ peak appears (figure 6(b)). The fits, with three parameters for one Gaussian peak (position, amplitude, FWHM), are performed without any constraint.

The different peak positions are displayed in table 1. Focusing on the Si-O absorption bands, a well-known peak under normal incident angle is observed at 1070 cm$^{-1}$ and attributed to the TO$_3$ stretching mode, which is also confirmed under Brewster angle at 1076 cm$^{-1}$ [47-51]. The LO$_3$ mode is present at 1222 cm$^{-1}$, compared with 1250 cm$^{-1}$ and 1254 cm$^{-1}$ in SiO$_2$ matrix respectively[47, 49] or 1227 cm$^{-1}$ and 1250 cm$^{-1}$ in the case of the SiO$_x$N$_y$ matrix[48, 52].

An important third peak is visible at 1128 cm$^{-1}$ under normal incidence which is blue-shifted at 1146 cm$^{-1}$ under Brewster angle. This peak cannot be ascribed to the TO$_4$ vibration band usually localized at 1200 cm$^{-1}$. This is confirmed with the fact that the appearance of the LO$_4$ mode under Breswter angle involves rather a red-shifted with a peak located at 1160-1170 cm$^{-1}$ [47, 53]. Such a peak is attributed to an "out-of-phase" stretching vibration mode of the Si-O bond. Indeed, Pai *et al.*, [50] based on previous results on local density of vibrational states (LDOVS) by Lucovsky *et al.* [54], underline a weak coupling of Si atom to O one causing this high frequency edge at 1150 cm$^{-1}$, the same position we noticed. This first result is confirmed by Kirk *et al.* who did an investigation of the disorder effect on IR absorption spectra in the a-SiO$_2$ host matrix and found that the interpretation of the IR spectra is somewhat in contradiction with an independent-oscillator model [47]. Indeed, by introducing disorder-induced mechanical coupling between the LO$_3$-TO$_3$ and LO$_4$-TO$_4$, a broad and significant vibration band appears in the range of 1076-1256 cm$^{-1}$, corresponding approximately to our bandwidth. Note that the amorphous character of our samples permits to detect a low contribution of the LO peaks even under normal incident angle as reported previously [55]. In addition, the blue-shift observed according to the appearance of the LO mode under Brewster angle for our samples, is in agreement with the spectra displayed by Kirk *et al.* [47]. This important shoulder in the 1100-1250 cm$^{-1}$ range under Brewster angle is also shown by previous works on silicon oxynitride host matrix deposited by PECVD technique[56-58].

The three last vibration bands are positioned at 867 cm$^{-1}$, 956 cm$^{-1}$ and 1024 cm$^{-1}$ under normal angle (or at 869 cm$^{-1}$, 953 cm$^{-1}$ and 1029 cm$^{-1}$ under Brewster angle) (see table 1). These energies correspond approximately to the positions ($\approx$903 cm$^{-1}$, 965 cm$^{-1}$, 1030 cm$^{-1}$) found by Naiman *et al.* after a dispersion analysis on (reoxidized) nitrided oxides on silicon [59], as well as the ones proposed by Moreno *et al.*[60]. The latter found modes peaking at 896 cm$^{-1}$, 962 cm$^{-1}$ and 1026 cm$^{-1}$ after performing modelling of silicon nitride system. But these peak positions are not ascribed to a specific vibration mode. Our first peak at 867 cm$^{-1}$ is well known and is originating from the Si-N stretching mode [52, 56, 61-64]. The difference between this first peak position at 867 cm$^{-1}$ and 903 cm$^{-1}$ (or 896 cm$^{-1}$) for the other authors, could suggest that we did not take into account the Si-O TO$_2$-LO$_2$ peaks at 810-820 cm$^{-1}$ (figures 3(a-b)) [47, 48].

As for the two last peaks at 956 cm$^{-1}$ and 1024 cm$^{-1}$, their attribution is more complicated. Focusing on the 956 cm$^{-1}$ peak, this band could be explained by the Si-OH bending mode but our matrix does not contain any hydrogen [65]. Another possibility could be the existence of an asymmetric mode Si-O-Tb bonds. But, as mentioned by Ono *et al.*[66], the heavier the element that binds with oxygen atom in Si-O is, the lower the vibrational frequency [66]. Then some authors have observed a vibrational frequency at about 900 cm$^{-1}$ for a Si-O-Pr [66] or, although Nd is heavier than Pr, around 910-950 cm$^{-1}$ for Si-O-Nd [67]. Yet, Tb atom is much heavier than the Pr or Nd ones and thus, a vibration at 955 cm$^{-1}$ for this bond could be excluded. Finally, such a peak at 956 cm$^{-1}$ was found in a 220 nm-thick SiO$_x$N$_y$ [68] and more precisely at the interface layer with the Si substrate whose thickness was about 2 nm on etched nitrided oxide layer [59]. In a detailed study of Ono *et al.*, this peak, positioned exactly at 960 cm$^{-1}$, has been attributed to the doubly bonded N atoms associated with two Si atoms ($\equiv$Si-N-Si$\equiv$) with the asymmetric stretching mode [69].

The last peak found at 1024 cm$^{-1}$ underlined by the dispersion analysis of Naiman *et al.* is ascribed, in their previous study, to the planar trigonal bonded nitrogen (Si$_3\equiv$N$_4$) [68]. Nevertheless, some studies assign this peak to a shoulder of the main (Si-N) stretching band at 870 cm$^{-1}$ [62, 70, 71].

| 5 Peak positions (cm$^{-1}$) Normal angle (figure 3(a)) | 6 Peak positions (cm$^{-1}$) Brewster angle (figure 3(b)) | Bonds attribution | Vibration modes | References |
|---|---|---|---|---|
| 867 | 869 | Si-N | TO stretching | [52, 56, 59-64] |
| 956 | 953 | ≡Si-N-Si≡ | TO Asymmetric stretching | [60, 69] |
| 1024 | 1029 | Si$_3$≡N$_4$ | TO Asymmetric stretching | [60, 68, 70] |
| 1070 | 1076 | Si-O | TO$_3$ (stretching) | [47-51] |
| 1128 | 1146 | Si-O | Out-of-phase (coupling TO$_3$-LO$_3$ and TO$_4$-LO$_4$) | [47, 49, 50] [56-58] |
| - | 1222 | Si-O | LO$_3$ | [47, 48, 52] |

Table 1: Peaks positions for normal and Brewster angles with their corresponding bond and vibration modes.

## 6.2. Variable Nitrogen excess

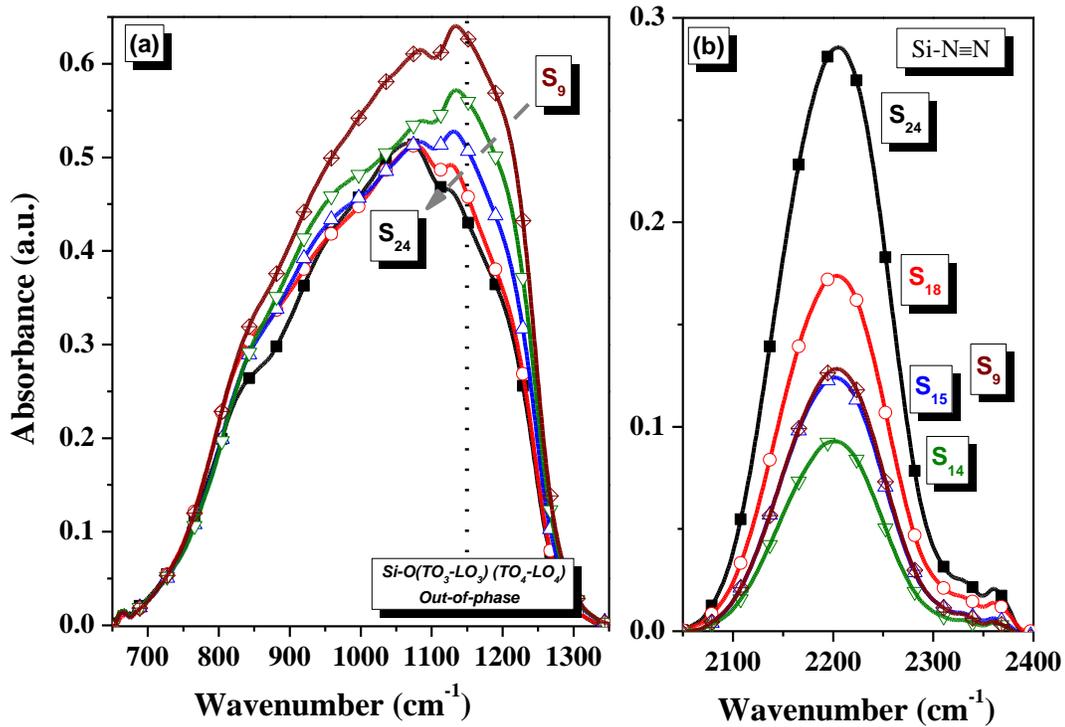

Figure 7. FTIR spectra recorded at Brewster angle (65 °) for AD films for different $N_{ex}$ values (a) in the 650-1350 cm$^{-1}$ range and (b) in 2050-2300 cm$^{-1}$ range. See also figure 11 in the supplementary data, which describes (a) the peaks positions, (b) FWHMs (reflecting the disorder) and (c) peak surfaces (number of bonds) coming from the fits analysis of figure 7(a).

The spectra in figure 7(a) show a blue-shift of the maximum intensity of about 60 cm$^{-1}$ towards 1075 cm$^{-1}$ with decreasing the $N_2$ flow inside the deposition chamber and so increasing $N_{ex}$. However, we note (see figure 11(a) in supplementary data) that during this $N_{ex}$ rise (9.4 % to 24.4 %), the different peaks positions blue-shifted slightly by only 20 cm$^{-1}$, except for the Si-N (TO) and Si-O (LO$_3$) modes which did not shift at all. Then this refocusing at 1075 cm$^{-1}$ is not due to a shift of all the peaks positions. Such a variation in shape is due to the decreasing of the out-of-phase peak intensity (see figure 11(b-c) in supplementary data). This decrease was already seen for AD films upon N incorporation [57] [58], as well as for annealed samples, [56] obtained by PECVD technique.

In the last 2050-2300 cm$^{-1}$ energy range (figure 7(b)), we note an absorption peak centered at 2202 cm$^{-1}$. Such a peak is originating from Si-H stretching [72] or isocyanate groups (N=C=O) or nitrile (C≡N) stretching bond [73]. But no H or C atoms were detected in our samples. Nevertheless, the intensities of these peaks seem to be sensitive to the $N_{ex}$ increase. Based on the study of adsorption of $N_2$ which is chemisorbed on Rh/SiO$_2$ sample, an important absorption peak at around 2200 cm$^{-1}$ at low temperature is explained by the N≡N bond linked to the Rh ion [74]. Consequently, our peak can be ascribed to the azide N≡N stretching mode [75], which gives a (Si-N≡N) resulting bond centered at 2202 cm$^{-1}$ depending of the $N_{ex}$ concentration. Such a peak attests also the N-rich character of our nitride matrix.

## 7. Emission properties

### 7.1. Photoluminescence properties

Figures 8(a) to 8(c) show the PL spectra of the samples annealed at different temperatures during one hour in a pure $N_2$ flow. These samples were excited at room temperature using a 325 nm line from a lamp source. This wavelength is non-resonant with energy levels of the Tb$^{3+}$ ion. As observed, each sample shows four PL peaks centered at about 490, 545, 590, and 624 nm, which correspond to the intra-4f $^5D_4 \rightarrow {}^7F_j$ (j = 6, 5, 4, 3) transitions of Tb$^{3+}$ ions, respectively as depicted in figure 8(a). These peaks have been already reported in previous studies on Tb:SiN layer obtained by reactive cosputtering [76] or PECVD deposition techniques [24, 25]. For each spectrum, the strongest peak intensity is located at 545 nm ($^5D_4 \rightarrow {}^7F_5$). This band has been selected to investigate the $T_A$ effect for various $N_{ex}$. The evolution of its integrated PL intensity with these parameters is shown in figure 8(b) as a function of $T_A$.

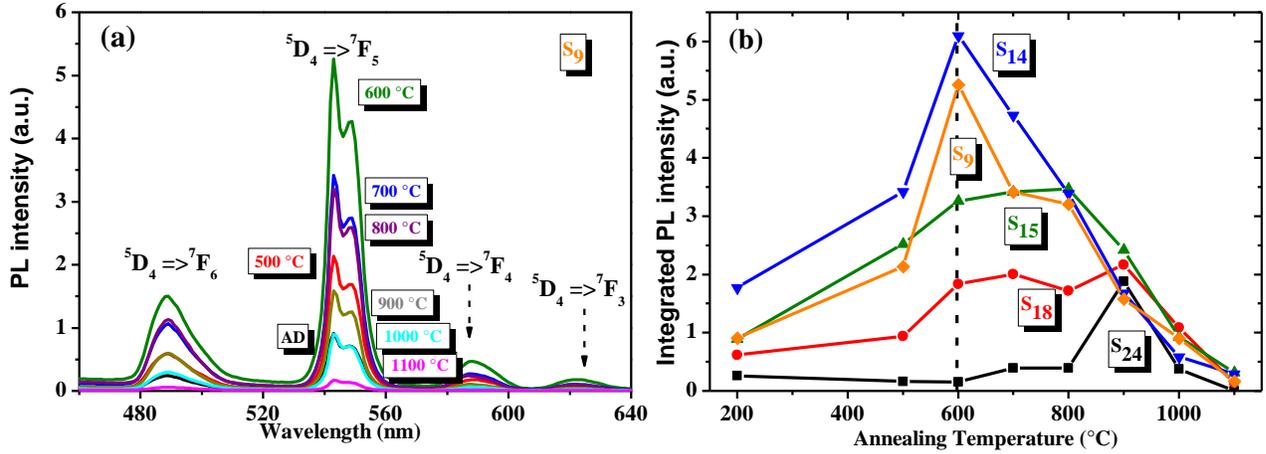

Figure 8. (a) PL spectra at 325 nm excitation wavelength of the $S_9$ sample annealed at different $T_A$ values during 1h (b) and the corresponding evolution of integrated PL intensity at 545 nm wavelength versus $T_A$ for all the samples.

For all samples, the PL intensity increases with $T_A$, reaches a maximum and starts to decrease for higher $T_A$ values. The former evolution of PL intensity is due to the passivation of non-radiative channels in the host matrix [77], whereas the latter could be assigned to the combination of two effects. Even if the porosities appear already in the AD sample (Figure 3(a)), the drop of the PL intensity could originate from the rising number of the porosities with annealing, offering more surface recombination centers and providing new non-radiative channels[78]. The second possible effect, which seems to be dominant, is the formation of Tb clusters, such as those observed in the TEM image in figure 5 at $T_A$=1200 °C and already detected at $T_A$=700 °C. These clusters have a detrimental effect on the PL intensity by the reduction of the number of optically active $Tb^{3+}$ ions [79]. Such effect is commonly seen for different rare earth-doped Si-based matrix as Nd:$SiN_x$ [80] or Er:SRSO [36].

It is interesting to note that the maximum PL intensity appears at different $T_A$ values depending on the $N_{ex}$. This optimum is reached at $T_A$=600 °C for samples $S_9$ and $S_{14}$ having both the lower $N_{ex}$ content (9.4 % and 14.7 %), while it appears at 800 °C for sample $S_{15}$ and 900 °C for samples $S_{18}$ and $S_{24}$ containing more $N_{ex}$. The temperatures corresponding to these maxima seem to be the limiting temperatures below which the cluster formation is not the predominant effect. Above this limit, the RE aggregates are driving down the PL intensity. Apparently, by introduction of a high N excess, the formation of clusters is delayed. This shift of the $T_A$ by almost 300 °C has already been observed for Er ions incorporated either in $SiO_2$ or $SiN_x$ [81]. In this study, the maximum intensity was underlined at $T_A$=800 °C for the silicate matrix while the maximum is not achieved for the $SiN_x$ matrix with a higher thermal budget. Our results confirm this point and show that the silicon nitride-based host matrix has a higher thermal threshold of RE clusterization than its silicon oxide counterpart [18].

*7.2. Time-resolved photoluminescence*

Although it is interesting to increase the $N_{ex}$ in the film in order to prevent the formation of Tb clusters, we have found that the maximum PL intensity is about 3 times lower with a high $N_{ex}$ value (figure 8(b)). Indeed, such a strongly non-stoichiometric composition may affect the quality of the host matrix and create many non-radiative defects. For deeper investigations, we conducted time-resolved PL measurements. Figure 9(a) displays the PL decays detected at 545 nm ($^5D_4 \rightarrow {}^7F_5$) wavelength as a function of $N_{ex}$, obtained for samples annealed at $T_A$=600 °C. The decay times deduced from these curves are shown in figure 9(b). Note that the lifetime of band tails-related PL (see §7.3) is usually in the nanoseconds scale which is too short to be measured by our experimental setup [82].

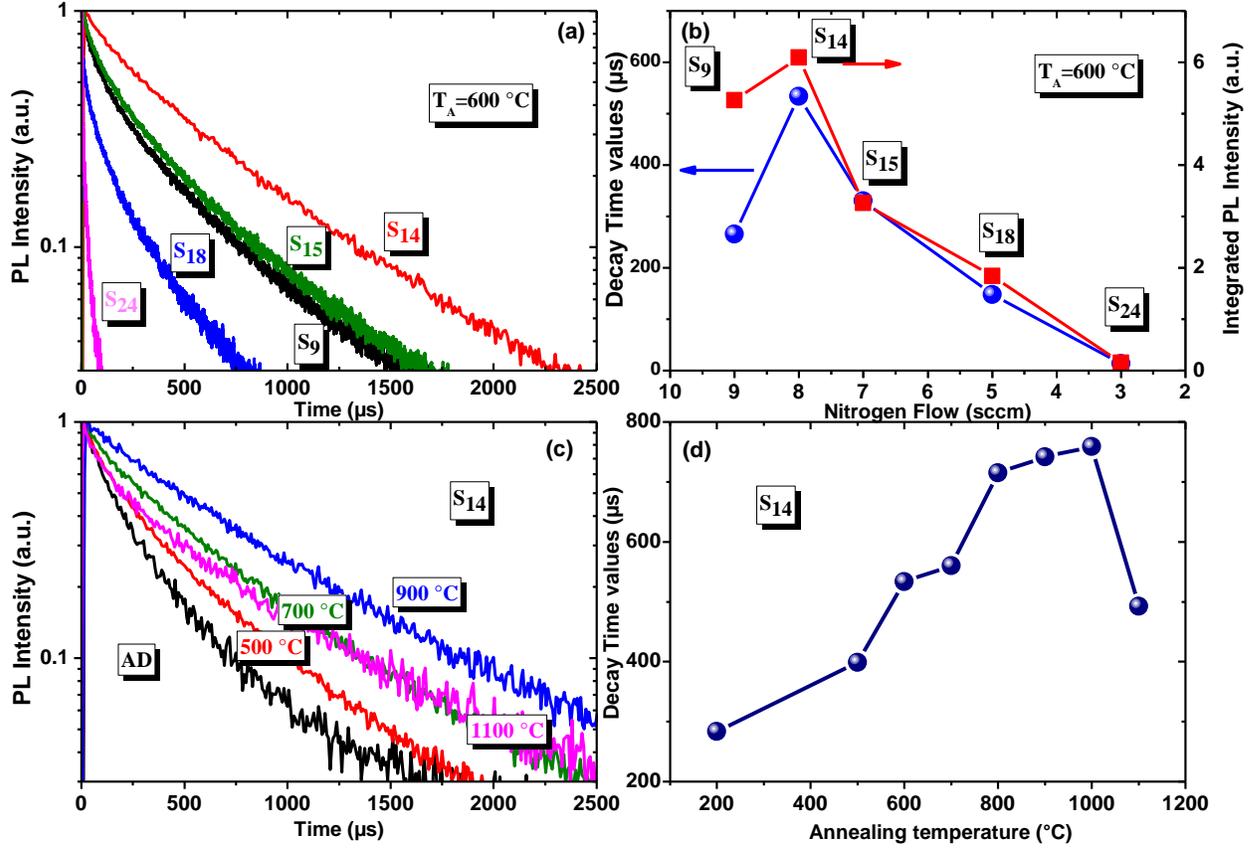

Figure 9. At the 545 nm ($^5D_4 \to {}^7F_5$) wavelength (a) the normalized decay curves for samples $S_9$ to $S_{24}$ annealed at $T_A$=600 °C and (b) their corresponding decay time with the evolution of the integrated PL intensity (c) the PL decay curves for the optimized layer $S_{14}$ annealed at the indicative $T_A$ and (d) their corresponding decay time.

The decay curves exhibit a non-exponential behavior and for this reason, we have calculated the mean decay time ($\tau_m$) using the following equation:

$$\tau_m = \int_0^\infty (I(t)/I_0)dt$$

where I(t) is a time dependent peak intensity while $I_0$ is the maximum intensity at t = 0s [83]. The lifetime obtained for the samples annealed at $T_A$=600 °C (figure 9(b)) for the lowest $N_{ex}$ amount is in the range of several-hundred-µs, comparable to the values reported by other authors [24]. The decay time increases slightly with $N_{ex}$ for $S_{14}$ sample and then a gradually decrease is noticed for higher $N_{ex}$ values. The integrated PL intensity displayed in figure 8(b) (dashed line at $T_A$=600 °C) for the same samples, and reported in figure 9(b) (right scale), shows a similar evolution. Therefore, the PL intensity behavior seems to be governed by the non-radiative recombination coming from the increasing population of defects with $N_{ex}$. This explains why the maximum PL intensity is about 3 times lower for the highest $N_{ex}$ (figure 8(b)).

To go further in the study of the annealing effect, we present the dependence of the PL decay time of the optimized sample $S_{14}$ detected at 545 nm with respect to $T_A$ (figures 9(c)-(d)). The PL lifetime increases with $T_A$ up to 1000 °C followed by a rapid decrease at 1100 °C. For this later, as explained above, this can be due to the large number of Tb-rich clusters formed leading to new non-radiative channels. Up to 1000 °C, a reduction of the non-radiative recombination takes place, due to the defects passivation, which causes the increase of the PL lifetime. In that case, one wonders why the PL intensity behavior has a maximum at $T_A$=600 °C (figure 8(b)), while the lifetime has a maximum shifted for 1000 °C (figure 9(d)). The long average lifetime measured (530 µs at $T_A$=600 °C) comes from the locally isolated $Tb^{3+}$ ions. The diffusion of $Tb^{3+}$ ions probably starts at 600 °C (Tb-rich clusters detected for the 700 °C annealed sample by TEM observations- not shown §5) reducing de facto the number of optically active $Tb^{3+}$ ions and concomitantly reduces the PL intensity. Consequently, between 600 °C and 1000 °C, the lifetime continues to increase due to the passivation of defects while we observe a reduction of the PL intensity. Such a behavior has been already detected on $Er^{3+}$ ions in silica glass by Polman *et al.* [84].

*7.3. Excitation mechanisms of $Tb^{3+}$ ions*

Figure 10(a) shows the PL spectra for both undoped and Tb-doped NRSON films annealed at 600 °C during 1h, measured using 380 nm (3.3 eV) excitation line. The undoped sample displays a broad band from about 2.5 to 3.2 eV (500 to 390 nm) originating from the recombination of excitons in band tails (BTs) states [26]. Indeed under such excitation wavelength, the BTs has a significant PL intensity compared to the excitation at 325 nm shown in figure 8(a) and described in our previous study [28]. The Tb-doped sample presents the features of the $Tb^{3+}$ ion peaks. The overlap between the $Tb^{3+}$ energy levels and the BTs is evident. This spectral overlap is very important since it allows indirect excitation of $Tb^{3+}$ ions via BTs states, leading to the observed $^5D_4 \rightarrow {}^7F_j$ transitions. The same behavior has already been observed for undoped [24, 25] and Tb-doped films[24] and is mainly due to the N dangling bond related recombination process.

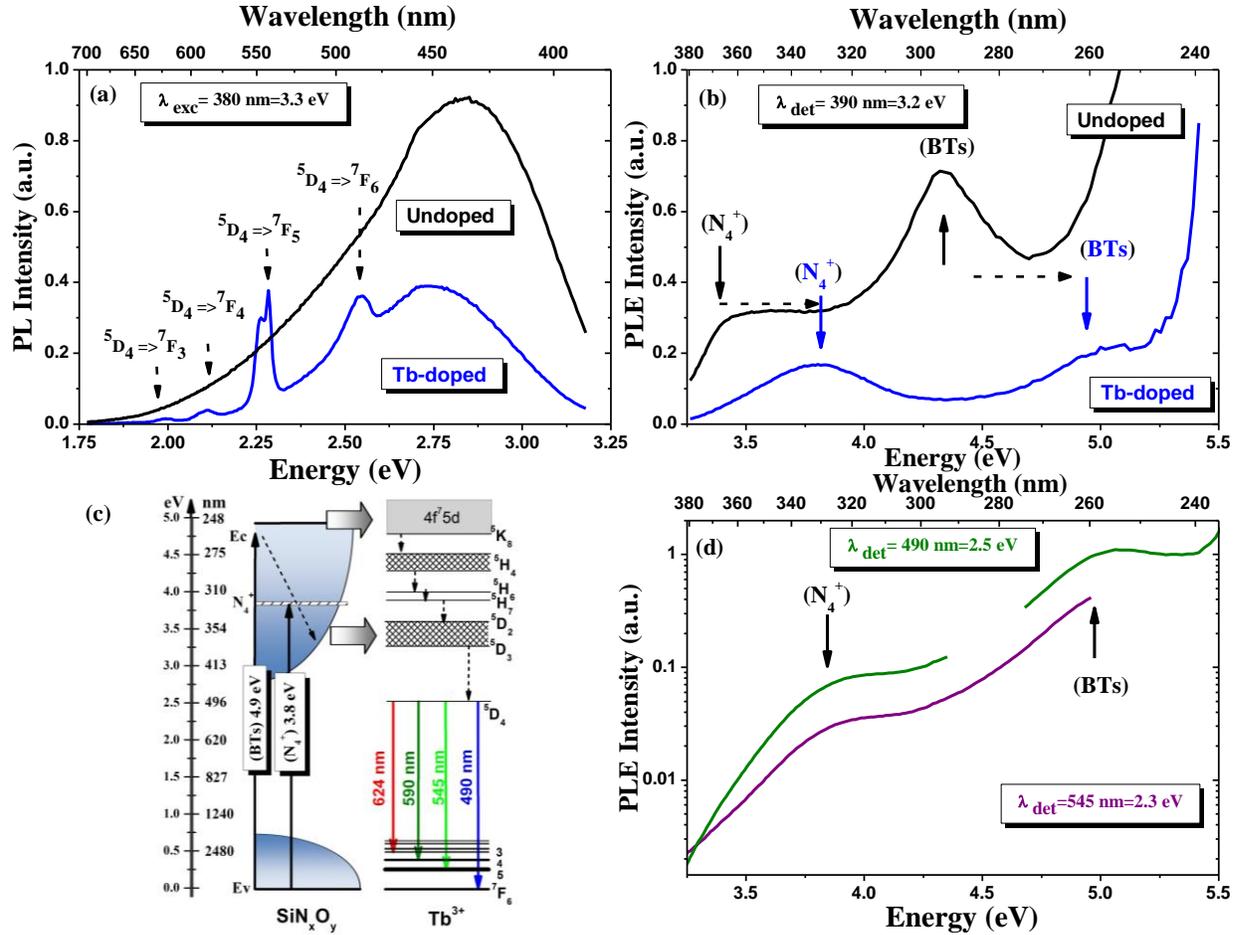

Figure 10. (a) PL spectra of undoped and Tb-doped films annealed at 600 °C during 1h (b) the associated PLE spectra detected at 390 nm (3.2 eV) related to the BTs (d) 490 nm (2.5 eV) ($^5D_4 \to {}^7F_6$) and 545 nm (2.3 eV) ($^5D_4 \to {}^7F_5$) related to the Tb$^{3+}$ transitions and (c) the schematic of the Tb$^{3+}$ energy diagram depicting the energy transfer between electronic energy levels.

Aiming at understanding the energy transfer mechanism to the Tb$^{3+}$ ions, PLE experiments were performed for both samples at 3.2 eV (390 nm) detection wavelength related to the BTs PL (Figure 10(b)) and also at 2.5 eV (490 nm) and 2.3 eV (545 nm), corresponding to the $^5D_4 \to {}^7F_6$ and $^5D_4 \to {}^7F_5$ transitions of the Tb$^{3+}$ ion (figure 10(d)). In the case of BTs detection, two peaks appear for the undoped film at 3.3 eV and 4.3 eV (figure 10(b) - black arrows). The first one is identified as a N defect state (N$_4^+$) [26], while the second is correlated to the absorption of the BTs just below the gap [28, 29]. It is linked to the Si–N bonds which introduce localized states due to N-2π lone pair electrons at the top of the valence band[85]. The recombination of holes and electrons in localized states associated with Si–N bonds explains the 2.5-3.2 eV peak in PL [77, 86, 87].

The Tb doping introduces some amount of oxygen, coming from the $Tb_4O_7$ chips, which results in increasing of the band gap. This leads to a blue shift of the two peaks of about 0.6 eV (dashed horizontal lines in figure 10(b)), to 3.9 eV and 4.9 eV, respectively. In any cases, the Tb doping may affect the BTs and its structure, because the peak intensities are not in the same ratio, in addition of their position changing. In the case of detection at $Tb^{3+}$ related peaks at 2.5 eV (490 nm) and 2.3 eV (545 nm) (figure 10(d)), both PLE spectra show an overall increase in the PL intensity with increasing the excitation energy. Above all, they have two peaks linked to the N defect state ($N_4^+$) and the absorption of the BTs underlined above, showing their active role in the $Tb^{3+}$ ions excitation.

A schematic of the different energy levels of the host matrix and the RE is suggested in figure 10(c). According to the previous studies [26, 28], the excitation of $Tb^{3+}$ in such thin films involves two types of excitations. First, when using high excitation energy photons (> 4.9 eV), the carriers-mediated excitation transfer occurs from the matrix to the $Tb^{3+}$ related states, such as the $^5K_8$ level or the excited electronic interconfiguration $4f^75d$. It is followed by a non-radiative relaxation in $Tb^{3+}$ ions towards the $^5D_4$ level. Second, when excitation energy is lower than the band gap of the matrix (< 4.9 eV), the excitons trapped in the BTs states or N defect states, can directly transfer their energy to $Tb^{3+}$ ions.

## 8. Conclusions

Tb-doped nitrogen rich silicon oxynitride (NRSON) films were deposited by co-sputtering method with variable $N_2$ flows and submitted to different thermal annealing treatments. The films were found to consist of two sublayers: a porous *top layer* and a bottom layer, so called *interface layer,* formed as a result of an unintended phase separation between $SiO_2$ and $Si_3N_4$. An explanation of this effect is proposed, involving spinodal decomposition. Upon a high annealing temperature at 1200°C, a formation of Tb-rich clusters is detected just below the separation between the interface and the top layer, revealing a higher thermal threshold of rare earth clusterization for the silicon oxynitride matrix in comparison to silicon oxide matrix.

Based on the RBS measurements, the $N_{ex}$ parameter has been introduced and found to vary in the opposite way than the N flow inside the deposition chamber. FTIR analysis with N excess showed several different peaks associated to the silicon oxide or nitride bonds. A classification of all the vibrational modes for such host matrix is proposed and compared to previous results in the literature. In particular, the "out-of-phase" stretching vibration mode of the Si-O bond at 1146 cm$^{-1}$, detected under Brewster angle, has been observed. We have found that its impact is reduced for high N excess values, i.e. for a low $N_2$ flow in the deposition chamber. Another peak linked to the azide N≡N bond at 2202 cm$^{-1}$ has also been identified.

The highest Tb$^{3+}$ PL intensity was obtained by optimizing the N incorporation and the annealing conditions. The PL spectra reveal the well-known peaks of Tb$^{3+}$ ion at 490 nm ($^5D_4 \rightarrow {}^7F_6$) and 545 nm ($^5D_4 \rightarrow {}^7F_5$) wavelengths. Finally, we found that the maximum PL intensity of the 545-nm Tb$^{3+}$ line corresponds to the annealed film at 600°C with 14.7% N excess, corresponding to the 8 sccm $N_2$ flow. The study in function of $T_A$ showed the role of Tb clustering in quenching the PL intensity. In particular, the analysis underlined that the effect of Tb clustering in silicon oxynitride matrix is significantly reduced, compared to the silicon oxide matrix, due to a higher thermal threshold of Tb clusterization. From this point of view and for the same thermal budget, the silicon oxynitride matrix allowed a much higher concentration of optically active Tb$^{3+}$ ions, which is beneficial for light emitting applications development.

The PL decay curves showed that the decay times of Tb$^{3+}$ ions are of the order of several-hundred-µs and revealed that the PL intensity was affected by the emergence of non-radiative defects with N excess. Besides, the values of the decay times continued to increase with the annealing temperature, while the PL intensity decreased, revealing the Tb-rich clusters formation, in agreement with the TEM observations. Finally, the mechanism of the excitation of Tb ions has been explored. PLE spectra of undoped and Tb-doped films have highlighted two main absorption peaks. The first peak was identified as N defect state ($N_4^+$), while the second was ascribed to the absorption of the band-tails just below the gap at 4.9 eV for the Tb$^{3+}$ doped sample. When excitation energy was lower than the band gap (< 4.9 eV), the excitons trapped in the band-tail states or N defect states, can directly transfer their energy to Tb$^{3+}$ ions.

**Acknowledgments**

This work was financially supported by the French Research National Agency through the GENESE project (No ANR-13-BS09-0020-01), the GENESIS EQUIPEX (ANR-11-EQPX-0020) and the LABEX EMC3 ASAP project. This work was also supported by the CEA/DSM/ENERGY contract (HOFELI Project) and the Chinese Scholarship Council (CSC) program and Polonium Partenariat Hubert Curien (PHC No 27720WC) Program.

The authors would like to thank Dr. Sophie Boudin from CRISMAT Laboratory (Caen, France) for the PL and PLE experiments.

In Poland, this work was funded by the National Science Centre in the framework of the Project No DEC-2012/05/D/ST7/01121.

# Structural and emission properties of Tb$^{3+}$-doped nitrogen-rich silicon oxynitride films

C. Labbé,[1] Y.-T. An,[1] G. Zatryb,[2] X. Portier,[1] A. Podhorodecki,[2] P. Marie [1], C. Frilay [1], J. Cardin [1] and F. Gourbilleau[1]

[1]CIMAP Normandie Univ, ENSICAEN, UNICAEN, CEA, CNRS, 6 Boulevard Maréchal Juin 14050 Caen Cedex 4, France

[2]Department of Experimental Physics, Wroclaw University of Technology, 50-370 Wroclaw, Poland

**Supplementary data**

**I) Explanations of the equation :** $\quad N_{ex}(\%) = \left( \dfrac{[N] - \frac{4}{3}[Si] + \frac{2}{3}[O] - \frac{7}{6}[Tb]}{[N]+[O]+[Si]+[Tb]} \right) \times 100$

The N excess displays the deviation from a perfect SiO$_2$ and Si$_3$N$_4$ phase mixture. It is calculated from the equation (1) as following:

The SiO$_x$N$_y$, depending on the ratio of x and y is composed of two phases SiO$_2$ (x=2, y=0) and Si$_3$N$_4$ (x=0, y=4/3)

$$SiO_xN_y \rightarrow \frac{x}{2} SiO_2 + \frac{y}{4} Si_3N_4$$

with $\quad \dfrac{x}{2} + \dfrac{3y}{4} = 1$

The Nitrogen excess was then compared to the stoichiometric Si$_3$N$_4$ matrix. We explain this calculus in two steps, without and with rare earth.

1) If we don't take account the rare earth's contribution to O.

$$N_{ex}(\%) = \left( \frac{[N]_{Tot} - [N]_{Si_3N_4}}{[N]_{Tot} + [O]_{Tot} + [Si]_{Tot}} \right) \times 100$$

where the term [element] represents the atomic concentration (at.%) of the aforesaid element. More specifically, the [element]$_{Tot}$ is the total amount of the element inside our films corresponding to the RBS measurements, while [element]$_{matrix}$ is the total amount of the element given by the aforementioned matrix.

$$\frac{[Si]_{Si_3N_4}}{[N]_{Si_3N_4}} = \frac{3}{4} \quad \rightarrow \quad [N]_{Si_3N_4} = \frac{4}{3} \times [Si]_{Si_3N_4}$$

and

$$\frac{[Si]_{SiO_2}}{[O]_{SiO_2}} = \frac{1}{2} \rightarrow [Si]_{SiO_2} = \frac{1}{2} \times [O]_{SiO_2}$$

due to the fact that the O comes from only SiO₂ matrix we can write :

$$[O]_{SiO_2} = [O]_{Tot}, \text{ and } [Si]_{SiO_2} = \frac{1}{2} \times [O]_{Tot}$$

then
$$[Si]_{Si_3N_4} = [Si]_{Tot} - [Si]_{SiO_2} = [Si]_{Tot} - \frac{1}{2}[O]_{Tot}$$

consequently

$$N_{ex}(\%) = \left(\frac{[N]_{Tot} - \frac{4}{3}[Si]_{Si_3N_4}}{[N]_{Tot} + [O]_{Tot} + [Si]_{Tot}}\right) \times 100 = \left(\frac{[N]_{Tot} - \frac{4}{3}[Si]_{Tot} + \frac{2}{3}[O]_{Tot}}{[N]_{Tot} + [O]_{Tot} + [Si]_{Tot}}\right) \times 100$$

2) If we take account the rare earth's contribution to O.

$$N_{ex}(\%) = \left(\frac{[N]_{Tot} - [N]_{Si_3N_4}}{[N]_{Tot} + [O]_{Tot} + [Si]_{Tot} + [Tb]_{Tot}}\right) \times 100$$

$$\frac{[Si]_{SiO_2}}{[O]_{SiO_2}} = \frac{1}{2} \rightarrow [Si]_{SiO_2} = \frac{1}{2} \times [O]_{SiO_2}$$

$$[Si]_{Si_3N_4} = [Si]_{Tot} - [Si]_{SiO_2} = [Si]_{Tot} - \frac{1}{2}[O]_{SiO_2}$$

$$\frac{[Tb]_{Tb_4O_7}}{[O]_{Tb_4O_7}} = \frac{4}{7} \rightarrow [O]_{Tb_4O_7} = \frac{7}{4} \times [Tb]_{Tb_4O_7}$$

$$[O]_{SiO_2} = [O]_{Tot} - [O]_{Tb_4O_7} = [O]_{Tot} - \frac{7}{4}[Tb]_{Tb_4O_7}$$

$$[Si]_{Si_3N_4} = [Si]_{Tot} - \frac{1}{2}[O]_{Tot} + \frac{7}{8}[Tb]_{Tb_4O_7}$$

due to the fact that the Tb comes from only Tb₄O₇ matrix we can assimilate: $[Tb]_{Tb_4O_7} = [Tb]_{Tot}$

$$[Si]_{Si_3N_4} = [Si]_{Tot} - \frac{1}{2}[O]_{Tot} + \frac{7}{8}[Tb]_{Tot}$$

Then $N_{ex}$ (%) is written:

$$N_{ex}(\%) = \left(\frac{[N]_{Tot} - \frac{4}{3}[Si]_{Tot} + \frac{2}{3}[O]_{Tot} - \frac{7}{6}[Tb]_{Tot}}{[N]_{Tot} + [O]_{Tot} + [Si]_{Tot} + [Tb]_{Tot}}\right) \times 100$$

To a rapid understanding in the publication we consider that: $[element]_{Tot} = [element]$

Then

$$N_{ex}(\%) = \left(\frac{[N] - \frac{4}{3}[Si] + \frac{2}{3}[O] - \frac{7}{6}[Tb]}{[N] + [O] + [Si] + [Tb]}\right) \times 100$$

## 2) *The elemental EFTEM map (O and N) on the $S_9$ sample annealed at 1200 °C*

The elemental EFTEM map (O and N) of the interface layer on the $S_9$ sample annealed at 1200 °C (not shown) has the same profile as that provided at 1000 °C (figure 4(d)). In contrast, for the top layer, we see the opposite of what we observed for a 1000 °C annealing treatment. Indeed, N has completely disappeared and the top layer is only composed of Si and O, where their ratio found by EDX spectra is close to the one of $SiO_2$ phase. Such oxygen amount does not seem to originate from the interface layer by diffusion due to the fact that its elemental map remains unchanged. Thus it could come from the degassing of the silica tube used in the furnace, at this specific high $T_A$ [1]. Surprisingly, this $SiO_2$ formation eliminates the porous columns to form exclusively the $SiO_2$.

## 3) *Additional figures of FTIR measurements: figure 11 linked to the figure 7(a)*

Figure 11: (a) the peak positions, (b) FWHM (reflecting the disorder) and (c) peak surface (number of bonds) coming from the fits analysis of the figure 7(a). The figures display also the corresponding error bar.

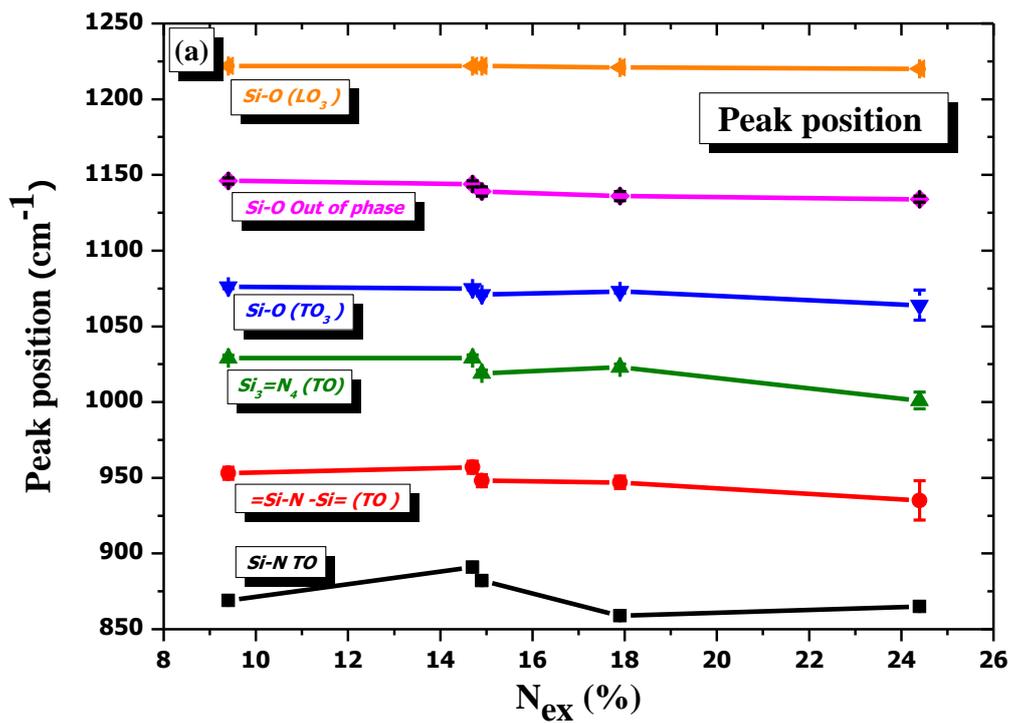

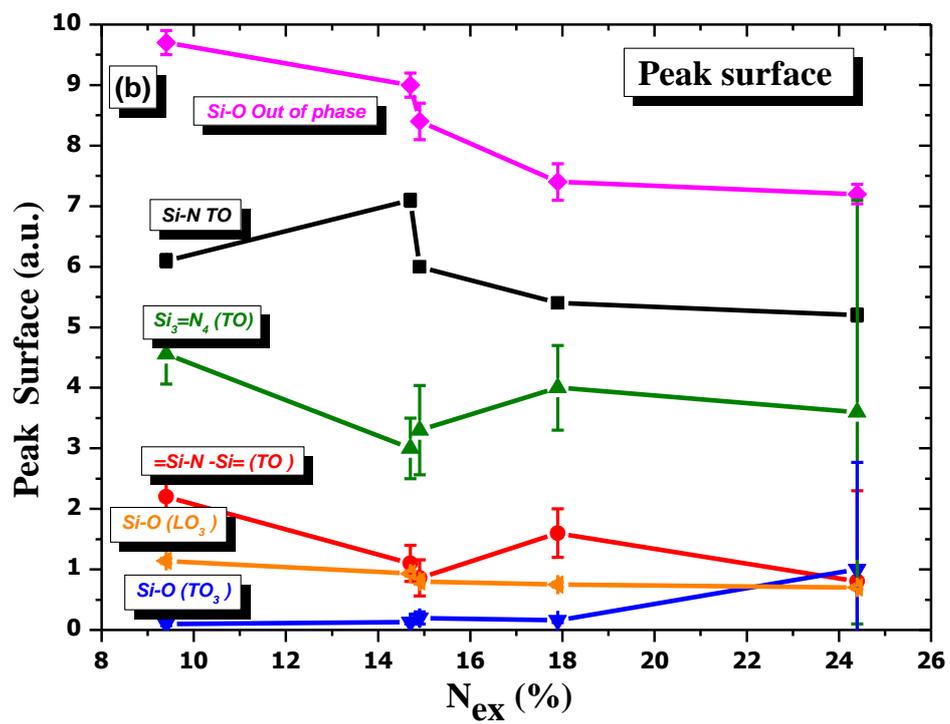

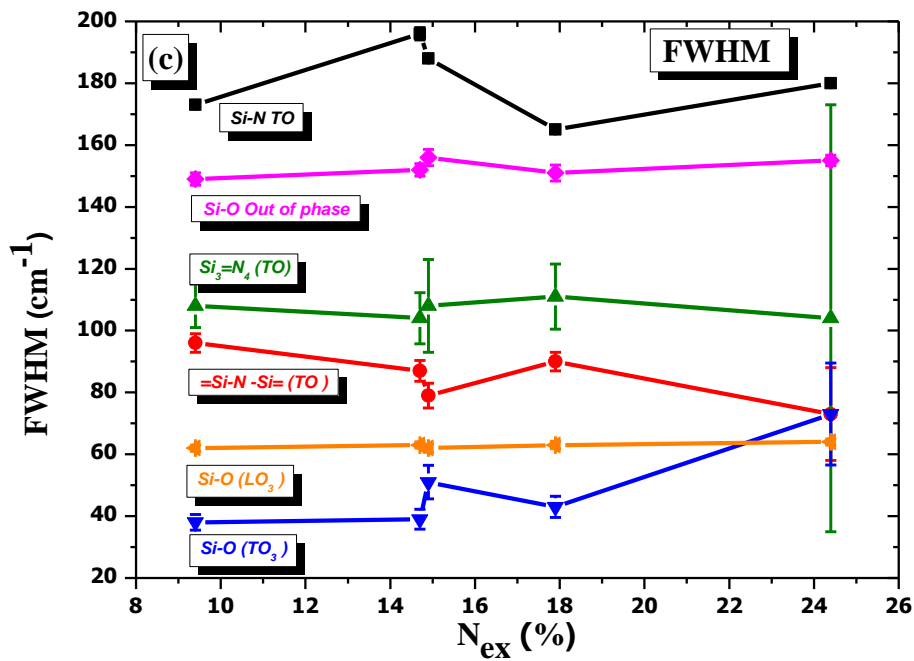

The out-of-phase peak intensity is linked to the introduction of disorder-induced mechanical coupling (Si-O) [2]. The Nex incorporation (diminution of Si) reduces the number of such bonds in a higher proportion than the other bonds, as shown in figure 11(b), but surprisingly no significant reduction of the FWHM of this phase is apparent (as shown figure 11(c)). This means that this disorder-induced mechanical coupling is the same, but with a reduction of its impact.